\documentclass[traditabstract]{aa}
\usepackage{epsfig,graphicx}
\usepackage{natbib}
\bibpunct{(}{)}{;}{a}{}{,}    

\begin{document}

\title{Scaling laws for magnetic fields on the quiet Sun}
\titlerunning{Scaling laws for magnetic fields on the quiet Sun}
\author{J.O. Stenflo\inst{1,2}}
\institute{Institute of Astronomy, ETH Zurich, CH-8093 Zurich \and Istituto Ricerche Solari Locarno, Via Patocchi, CH-6605 Locarno Monti, Switzerland}

\date{}

\abstract{
The Sun's magnetic field is structured over a range of scales that
span approximately seven orders of magnitudes, four of which lie
beyond the resolving power of current telescopes. Here we have used a
Hinode SOT/SP deep mode data set for the quiet-sun disk center in
combination with constraints from the Hanle effect to derive scaling
laws that describe how the magnetic structuring varies from the
resolved scales down to the magnetic diffusion limit, where the field
ceases to be frozen-in. The focus of the analysis is a derivation of
the magnetic energy spectrum, but we also discuss the scale dependence
of the probability density function (PDF) for the flux densities and
the role of the cancellation function for the average unsigned flux
density. Analysis of the Hinode data set with the line-ratio method
reveals a collapsed flux population in the form of flux
tubes with a size distribution that is peaked in the 10-100\,km
range. Magnetic energy is injected into this scale range by the instability mechanism of 
flux tube collapse, which is driven by the external gas pressure in 
the superadiabatic region at the top of the convection zone. This
elevates the magnetic energy spectrum just beyond the telescope resolution 
limit. Flux tube decay feeds 
an inertial range that cascades down the scale spectrum to the magnetic diffusion limit,
and which contains the tangled, ``hidden'' flux that is known to exist
from observations of the Hanle effect. The observational constraints 
demand that the total magnetic energy in the hidden flux must be of the same
order as the total energy in the kG flux tubes. Both the flux tubes
and the hidden flux are found to be preferentially located in the intergranular
lanes, which is to be expected since they are physically related. 
}

\maketitle

\section{Introduction}\label{sec:intro}
Different types of scaling laws have been used to explore how solar
magnetic fields vary with scale size. Since the structuring is
produced by turbulent convection and turbulent properties are usually
described in terms of energy spectra, there have been a number of
attempts to derive and analyse such spectra for magnetic
fields observed in solar magnetograms 
\citep{stenflo-nakagawa73,stenflo-nakagawa74,stenflo-knobloch81,stenflo-knoblochrosner81,stenflo-petrovay01,stenflo-abramenko01,stenflo-abramenko10,stenflo-abramenko11}.
The obtained spectra refer to the directly observed domain above the
resolution cut-off of the data sets used. However, since the physical
cut-off, the magnetic diffusion limit, where the magnetic field ceases
to be frozen-in, is located at scales that are about four orders of
magnitude smaller, the observationally resolved domain covers only the
large-scale end of the spectrum.  With the assumption that the
  fundamental dissipation of the magnetic field occurs as ohmic
  dissipation, the diffusion limit is the scale at which the magnetic
  Reynolds number becomes unity, or in other words, where the ohmic diffusion
  time scale becomes comparable to the dynamic time scale 
  \citep[cf.][]{stenflo-dewijn09}. 

Since the resolved magnetic fields appear to exhibit a high degree of
self-similarity and scale invariance, they have been explored by
fractal analysis
\citep[e.g.][]{stenflo-lawetal93,stenflo-lawetal96,stenflo-cadetal94,stenflo-janssen03,stenflo-abramenko10f}. 
Another useful concept related to the multi-fractal nature of the
field is the so-called cancellation function, introduced and
applied by
\citet{stenflo-pietarila09} in the analysis of quiet-sun data from
the Hinode SOT/SP instrument
\citep{stenflo-kosugi07,stenflo-suematsu08,stenflo-tsuneta08}, leading
to the conclusion that at least 80\,\%\ of the 
magnetic flux is invisible at the Hinode resolution due to
cancellation of the contributions from the opposite magnetic
polarities within the spatial resolution element. This confirms the
long-standing (three decades old) conclusion from observations of the Hanle effect
\citep{stenflo-s82,stenflo-s87,stenflo-trujetal04} that the photosphere is
seething with an ocean of tangled, ``hidden'' magnetic flux that is
invisible to Zeeman-effect observations with the available angular
resolution. 

The statistical properties of the magnetic fields need to be described
in terms of distribution functions, the most common of which is the
probability density function, PDF, of the magnetic flux densities or
of the field strengths. To avoid confusion we here want to reserve the
term ``field strength'' for the resolved, unsmeared field, and use the
term ``flux density'' for the field that has been smoothed by the
spatial resolution window. As no quiet-sun field structures are
resolved with the Hinode resolution, the directly observed PDF, of the flux
densities, is very different from the PDF for the field strengths,
which can only be inferred by indirect methods. The flux density PDF
for the quiet Sun, as observed with Hinode, is characterized by an
extremely narrow core that is centered at zero field and can be
approximated by a stretched exponential, and by extended wings that
decline quadratically and extend out to the kG region
\citep{stenflo-s10aa}. 

A powerful method to infer the magnetic structure at scales beyond
the spatial resolution limit is the Stokes $V$ line-ratio technique,
which was introduced and used for the discovery four decades ago \citep{stenflo-s73}
that a large fraction of the 
quiet-sun magnetic flux is in the form of strong-field (kG)
flux tubes with small filling factors (of order
1\,\%\ in quiet regions). A theoretical explanation for this extreme
intermittency was provided by the instability mechanism of flux tube
collapse 
\citep{stenflo-parker78,stenflo-spruit79,stenflo-sprzw79,stenflo-unnoando79}. 
Application of the line-ratio technique to Hinode 
SOT/SP data has allowed the two distinct flux populations, the
collapsed and uncollapsed population, to be identified and
statistically separated \citep{stenflo-s10aa}. In a follow-up work
\citep{stenflo-s11aa} the Hinode line-ratio data could be used to
derive a histogram for the flux tube sizes (which is peaked in the
10-70\,km range), and the collapsed population was found to be
preferentially located in the intergranular lanes. 

While it has been known since the early 1970s that much of the total
quiet-sun magnetic flux is in collapsed form and since the early 1980s
that a large fraction of the flux must also be in hidden form (invisible in
magnetograms), a comprehensive physical picture that connects these two
apparently disjunct aspects of solar magnetism has been missing. The
present work can be seen as an attempt to fill this gap.

Often the term ``local dynamo'' is used in discussions of
  small-scale magnetic structuring on the Sun. Here we avoid such
  terminology, since we find no evidence that the small-scale
  structuring that we will be discussing is produced by turbulent
  amplification of a weak seed field. Instead, as we will see, there
  is evidence that the very substantial amount of magnetic energy that
  exists on small scales (below a few km) is physically related to and
  fed from the magnetic energy at large scales, which is generated by
  the global dynamo. This implies that the rate of magnetic energy
  dissipation is faster than the rate of energy production by local
  dynamo action. 

Our analysis of the magnetic scaling laws is based on the
same Hinode SOT/SP data set for the quiet-sun disk center that we used
before \citep{stenflo-s10aa,stenflo-s11aa}. This time the main
focus is on the magnetic energy spectrum, its determination in the
resolved domain, and its continuation throughout the unresolved domain
down to the magnetic diffusion limit. In particular we explore the
energy that is injected into the spectrum by the mechanism of flux
tube collapse and try to identify the spectral location and the energy
contents of the hidden magnetic flux that is responsible for the
observed depolarization due to the Hanle effect. We further derive the
PDF for the intrinsic field strengths that is needed to satisfy the
joint Hanle and line-ratio constraints and discuss the nature and role
of the cancellation function.

\section{Energy spectra: concepts and definitions}\label{sec:concepts}
The determination of the spectral energy density from solar
observations of magnetic and velocity fields might seem to be a
straightforward matter: compute the Fourier transform and square its
absolute value to obtain the power spectrum. However, there are
different ways in which the energy spectra can be defined, which
sometimes leads to confusion, and the spectra are affected by noise
and observational cut-offs. In addition there are technical issues
like apodization and interpolation techniques. 

\subsection{Relations between the four versions of the energy spectra}\label{sec:4versions}
In a 2-D image the spatial $x$ and $y$ coordinates have their
counterparts in the spectral domain in the wave numbers
$k_x=2\pi/\Delta x$
and  $k_y=2\pi/\Delta y$, with total wave number 
$k=\sqrt{k_x^2+k_y^2}$. We can distinguish between four versions of the
spectral energy density: the 1-D power spectra in the $x$ and $y$
directions, which after normalization (described below) we denote by 
$E_x(k_x)$ and $E_y(k_y)$, the 1-D spectrum in $k$ space, which we
denote $E_{1D}(k)$, and the 2-D spectrum $E_{2D} (k_x,k_y)$, which in the
axially symmetric case can be written as $E_{2D} (k)$. 

The normalization condition is 
\begin{eqnarray}
\int_{-\infty}^{+\infty} \! E_{x,y}(k_{x,y})\,{\rm d}k_{x,y}\, &\,= {\cal
  E}\,,\nonumber \\ \int_0^\infty \! E_{1D} (k)\,{\rm d}k \, &\,= {\cal
  E}\,, \\ \int E_{2D} (k_x,k_y) \,{\rm d}k_x\,{\rm d}k_y \, &\,= {\cal
  E}\,,\nonumber\label{eq:norm}
\end{eqnarray}
where ${\cal  E}$ is the average energy density of the data set. 

In SI units we have 
\begin{equation}
{\cal  E}=\langle B^2\rangle\,/\,(2\mu_0)\label{eq:magnorm}
\end{equation}
for the magnetic field, where $\langle B^2\rangle$ is the spatial
average of $B^2$, and 
\begin{equation}
{\cal  E}={\textstyle{1\over 2}}\rho\,\langle v^2\rangle\,,\label{eq:velnorm}
\end{equation}
where $v$ is the velocity vector, and $\rho$ is the mass density in
the line-forming layer of the 
solar atmosphere, needed for the comparison between the magnetic and
kinetic energy densities with respect to the issue of equipartition. 

In the case of axial symmetry for $E_{2D} (k_x,k_y)$, which can safely
be assumed for the disk center of the quiet Sun, the spectrum is
independent of azimuthal angle, which means that 
\begin{equation}
\int E_{2D} (k_x,k_y) \,{\rm d}k_x\,{\rm d}k_y \,=2\pi\!\int_0^\infty\! 
k\,E_{2D} (k)\,{\rm d}k \,,\label{eq:axisym}
\end{equation}
from which follows that 
\begin{equation}
E_{1D} (k)\,=2\pi\,k\,E_{2D} (k)\,.\label{eq:12rel}
\end{equation}

In the axisymmetric case there is also a direct relation between
$E_{x,y}(k_{x,y})$ and $E_{2D} (k)$ via the Abel transform. If we for
instance know $E_y(k_y)$, we obtain $E_{2D} (k)$ from the inverse
Abel transform \citep[introduced in solar physics for the analysis of solar
granulation spectra by][]{stenflo-uberoi55} through 
\begin{equation}
E_{2D} (k)\,=\,-{1\over 2\pi}\,\int_k^\infty\,{{\rm
    d}E_y(k^\prime)\over{\rm d}k^\prime}\,{{\rm
    d}k^\prime\over\sqrt{{k^\prime}^2 -k^2}}\,.\label{eq:abel}
\end{equation}
Together with Eq.~(\ref{eq:12rel}) 
we are then in a position to convert
$E_y(k_y)$ into $E_{1D} (k)$. 

If $E_y(k_y)$ is given by a power law, then $E_{1D} (k)$ is also
given by a power law with the same exponent. In contrast, the 2-D
spectrum $E_{2D} (k)$ does not obey the same power law. 

We note that it is the version $E_{1D} (k)$ of the energy spectrum
that is used in turbulence theory, for instance when the $-5/3$ power law is
referred to in \citet{stenflo-kolmogorov41} theory.

\subsection{Upper and lower cut-offs}\label{sec:cutoff}
The ``true'' 2-D solar image is smeared by the effective resolution window
(often called point spread function), determined by the size and
quality of the telescope (and by atmospheric seeing in the case of ground-based
instruments). This spatial smearing represents a convolution in the
$x$-$y$ plane, while in the $k_x$-$k_y$ plane the power spectrum of
the image gets multiplied by the 2-D power spectrum of the resolution window,
the modulation transfer function (MTF). The MTF is unity for small
wave numbers $k$, but drops off steeply towards zero in the vicinity
of the $k$ that represents the resolution limit. This cut-off defines the upper
boundary in $k$ space of the observationally determined spectrum. 

According to the sampling theorem, there have to be at least two
pixels per resolution element (to avoid aliasing). If the pixel size
is $p$, then the upper limit in $k$ space is $2\pi/(2p)$, which
approximately coincides with the resolution limit when one chooses two
samples per resolution element.  

There is a corresponding cut-off at low wave numbers, determined by
the maximum size $L$ of the field of view over which the energy spectrum
is determined: $2\pi/L$.

\subsection{Field strengths and flux densities}\label{sec:bbar}
The observed magnetic and velocity fields are not the ``true'' fields,
but represent spatially smeared quantities due to the finite
resolution of the instrument. To make this distinction explicit, we
will mark the smeared (observed) quantities with a bar above the
symbol. Thus the true velocity field along the line of sight (the $z$
axis) is $v_z$, while the corresponding observed quantity is ${\bar
  v}_z$. Similarly, for the magnetic field we distinguish between
$B_z$ and ${\bar B}_z$. 

To avoid confusion, which has been abundant in previous literature, we
employ the terminology introduced in \citet{stenflo-s10aa}, and never
use the term ``field strength'' for ${\bar B}$ but the term 
``flux density'' instead.  We reserve the term ``field strength'' exclusively 
for $B$, the field seen with infinite resolution (a theoretical limit that is
unreachable in practice). On the quiet Sun no flux elements are spatially resolved
with the Hinode resolution. The quantities $B_z$ and ${\bar B}_z$
therefore differ profoundly, as we will see explicitly and
quantitatively later when comparing the
probability density functions for the field strengths and the flux
densities. 

The set of equations given in Sect.~\ref{sec:4versions} formally
represent the unsmeared quantites. To obtain them in terms of the smeared
quantities we place a bar above every physical quantity. If we
denote the effective MTF of the telescope system by $T(k)$, 
the relation between the apparent energy spectrum ${\bar E}_{1D}(k)$
based on the smeared quantities, and the ``intrinsic'' energy spectrum
$E _{1D}(k)$, becomes 
\begin{equation}
{\bar E}_{1D}(k)\,=\,E_{1D} (k)\,T(k)\,.\label{eq:etk}
\end{equation}

\subsection{Apodization and interpolation}\label{sec:apod}
The Fourier integrals that are used for the determination of the
spectrum extend 
to infinity, while the available field of view is
finite. Mathematically the finite field of view can be described in terms of truncation of
an ideally infinite image through multiplication by a window function
that is unity within the field of view and zero outside. Due to this
truncation the ``true'' Fourier spectrum gets
convolved (or smeared) by a function that is the Fourier
transform of the field-of-view window function. 

As the Fourier transform of a rectangular window with sharp edges has
large side lobes which could introduce unwanted spurious effects in
the spectrum, one usually makes the sharp edges smooth by 
apodization. In the present work we apodize by tapering off the window
function over the outer 1/3 of the field of view with a cosine bell function. 

Since the $k$ increment in the discrete, numerical representation of
the energy spectrum becomes very course as we go to smaller wave
numbers, it is desirable to make the representation and plotted curves
smoother by interpolation. The standard way of doing this is not by direct
interpolation in $k$ space (which would not work well), but by
extending the formal field of view with zeros, and then rescaling the
resulting energy spectrum by dividing the spectrum with the fraction
of the field of view that is occupied by real data (the ``data filling
factor''). In our case we have extended the length of our field of
view by a factor of 6 to get a 6 times finer $k$ grid.

\subsection{Influence of noise and choice of data set}\label{sec:noise}
Since measurement noise can seriously distort the energy spectrum, it
is of great importance to use data with the best possible S/N ratio,
and in addition test the sensitivity of the data set to added
artificial noise. 

The best-quality magnetic-field data presently available in terms of
well-defined high spatial resolution combined with low noise has been obtained with
the Hinode satellite. There are two Hinode instruments that provide
two types of magnetic-field data: the filter
magnetograms in the Na\,{\sc i} D$_1$ line (FG data), and the
spectrograph Stokes line profile data of the Fe\,{\sc i} 6301.5 and
6302.5\,\AA\ lines (SOT/SP data). As the FG data are by far noisier
than the SOT/SP data and in addition provide no information on the
vector field or the filling factors, they are not suited for the
present analysis. 

Each exposure with the SOT/SP instrument provides 1-D spatial
information of the magnetic structuring along the spectrograph slit,
here defined as the $y$ direction (along the heliographic N-S
direction). 2-D images of the magnetic field distribution in the
$x$-$y$ plane can be built up by step-wise scanning in the $x$
direction, but since the time it takes to cover a significant $x$
range is generally much larger than the evolutionary time scale of the
magnetic elements, the 2-D images do not represent snapshots but mix
spatial structuring with evolutionary effects. In
addition, to cover a reasonable field of view in the $x$ direction, the exposure time for
each frame needs to be relatively short, with the consequence of more noise in
these 2-D magnetograms. Therefore they do not represent the best
data set choice for a spectral analysis of quiet-sun magnetic
fields, for which the observable polarization signatures are very weak. The situation
is different for active-region analysis, where the signals are
strong. 

Since low noise is so essential for quiet-sun analysis, and since we
need to avoid confusion between spatial structuring and evolutionary
effects, the clearly best choice of data set is the SOT/SP deep mode
recording with the slit at a fixed position at the quiet-sun disk
center. Our deep mode data set consists of a time series of 727
exposures, each with an 
integration time of 9.6\,s, obtained on February 27, 2007. It is
the identical data set that has already been subject to in-depth
analysis by \citet{stenflo-litesetal08} and
\citet{stenflo-s10aa,stenflo-s11aa}. It provides high-quality
magnetic-field data for the 1024 pixels of size 0.16\,arcsec along the $y$ direction. 

The Stokes profile data for each pixel of this data set have been
converted to vertical flux densities ${\bar B}_z(y)$ and total flux
densities (the magnitude of the flux density vector) ${\bar B}(y)$ as
described in \citet{stenflo-s10aa}, making use of the 6302/6301
line-ratio information to account for the non-linear effects on the
pixel-averaged flux densities due to the 
subpixel flux tube structuring and line weakenings. Since the noise in
the transverse field 
(after conversion from fractional polarization to G) is much larger
(of order 25 times) than the noise in the longitudinal field, the determined
total flux densities are much noisier than the vertical flux
densities. 

\begin{figure}
\resizebox{\hsize}{!}{\includegraphics{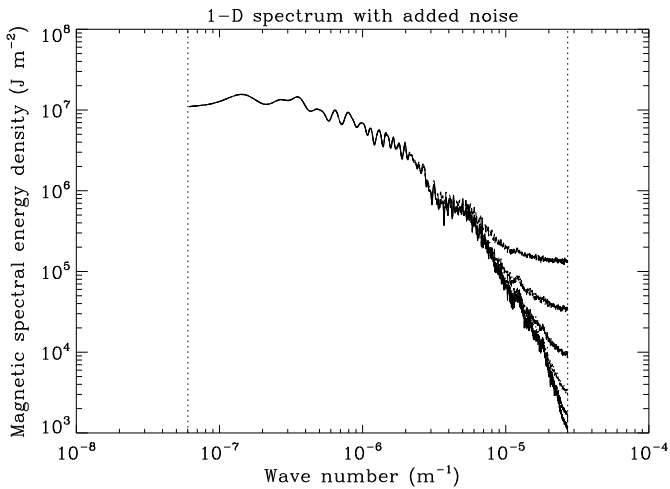}}
\caption{Magnetic energy spectra ${\bar E}_y(k_y)$ of the vertical flux densities, without and
  with various amounts of random noise added to the original data, 
  for which the measurement noise $\sigma_{\rm obs}$ is 1.4\,G. The five curves
  that deviate increasingly from the original spectrum for the largest
  wave numbers represent added noise with standard deviations of 1, 2,
  4, 8, and 16 $\times \sigma_{\rm obs}$. The dotted lines mark the
  boundaries of the observed spectrum, determined to the right by the
  Hinode resolution (0.232\,Mm), to the left by the effective slit
  length (104\,Mm). 
}\label{fig:addnoise}
\end{figure}

In Fig.~\ref{fig:addnoise} we show the computed 1-D spectral energy
density ${\bar E}_y(k_y)$ for the vertical flux densities, obtained by
averaging the 727 power spectra along the slit (after apodization
etc. as described in Sect.~\ref{sec:apod}). The observational noise in
the ${\bar B}_z$ data is $\sigma_{\rm obs}=1.4$\,G. To explore the effect of more noise, we have
overplotted spectra for which random noise with a standard deviation
of 1, 2, 4, 8, and 16 $\times \sigma_{\rm obs}$ has been added to the
${\bar B}_z(y)$ values. We see that adding only one $\sigma_{\rm obs}$
has almost no discernible effect on the spectrum, which confirms that
the original spectrum is not significantly affected by noise. However,
as more noise is added, the originally steep spectrum gets raised to
become flatter (since the noise contribution is spectrally
flat). As expected it is the largest wave numbers that are
most affected by noise.

\section{Equipartition between kinetic and magnetic energy}\label{sec:equipart}

\begin{figure}
\resizebox{\hsize}{!}{\includegraphics{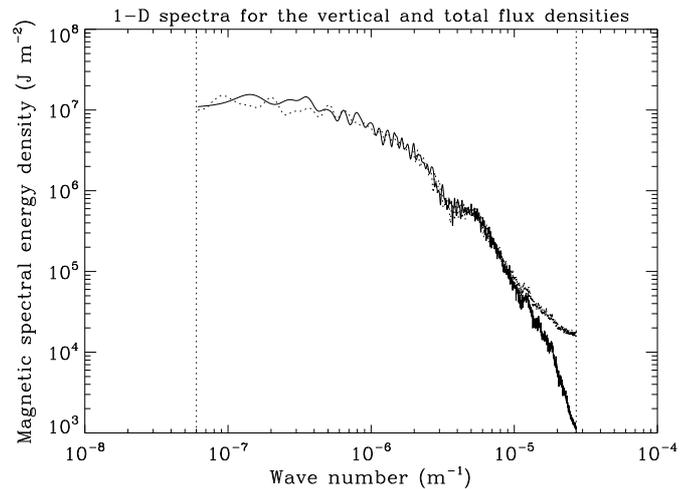}}
\caption{Magnetic energy spectra ${\bar E}_y(k_y)$ for the vertical (solid line) and
  total (dotted line) flux densities. The solid line is the same as
  the lowest curve in Fig.~\ref{fig:addnoise}. The dotted curve agrees
  with the solid one 
  except at the largest wave numbers, but the deviation there is most
  likely due to noise (as the transverse Zeeman effect that is used to
  derive the total flux densities introduces non-Gaussian noise that
  is much larger than the $\sigma_{\rm obs}$ for the vertical flux
  densities). 
}\label{fig:bvbsp}
\end{figure}

In Fig.~\ref{fig:bvbsp} we compare (without added noise) the 1-D
spectra of the vertical and total flux densities. The
two spectra are nearly indistinguishable except for the largest wave
numbers, where the spectrum for the total flux densities flattens
out. However, when comparing with the previous Fig.~\ref{fig:addnoise}
we notice that the shape of this flattening is the same as obtained
when random noise is added. Since we know that the noise level for the
total flux densities is much larger than for the vertical flux
densities, this behavior is expected. 

There is therefore no evidence for a significant deviation between the
spectra for the total and the vertical flux densities. This may seem
surprising, because if for instance the field vectors would have an
isotropic angular distribution, then the spectral density for the
total flux density would be larger than that of the vertical flux
density by a factor of 3 (due to the 3 spatial degrees of
freedom). However, as was found in \citet{stenflo-s10aa}, the angular
distribution of the field vectors is peaked around the vertical
direction, except for the smallest flux densities, where the
distribution becomes nearly isotropic (although the smallest flux
densities are the ones that are most affected by noise). The
contributions to the spectral energy density are proportional to ${\bar B}^2$,
which favors the contributions from the largest flux densities, which
are the most vertical. 

Therefore we will in the following take the energy spectrum for the
vertical flux densities, being virtually noise free, to also represent
the spectrum for the total flux densities. In the case of the velocity
spectrum, however, the assumption of an isotropic angular distribution
should be good (with some question marks for the lowest wave numbers
or largest scales). Therefore we need to multiply the energy spectrum
for the observed vertical velocities ${\bar v}_z$ by 3, and in
addition attach the factor ${\textstyle 1\over 2}\rho$ to obtain the
kinetic energy spectrum that can be compared with the magnetic energy
spectrum. The value of the mass density $\rho$ does not come from
observations, but must be chosen from a model atmosphere. The relevant
choice for comparison with the magnetic energy density is the $\rho$
that represents the height of formation of the Fe\,{\sc i} lines on
which the analysis is based. 

\begin{figure}
\resizebox{\hsize}{!}{\includegraphics{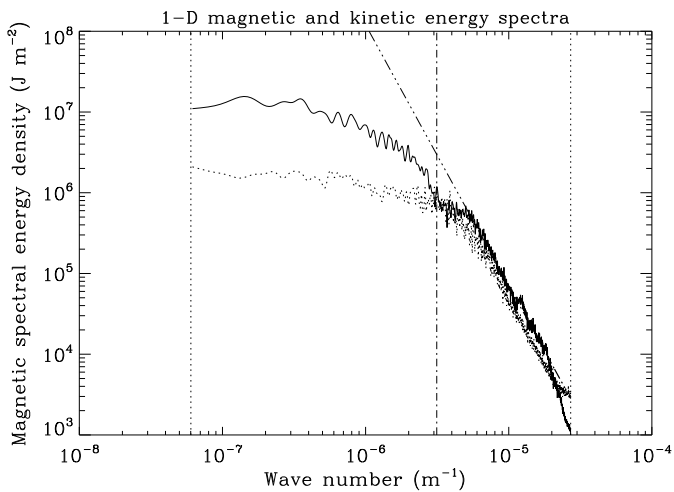}}
\caption{Comparison between the magnetic (solid) and kinetic (dotted)
  energy spectra ${\bar E}_y(k_y)$. The slanted dash-triple-dot line
  represents a power law with exponent $-3.23$. The vertical dotted
  lines mark the boundaries of the observational domain, while the
  vertical dash-dotted line marks the 2\,Mm scale, for
  reference. There is approximate equipartition between the magnetic
  and kinetic energy densities over the range 0.3-2.0\,Mm, and
  both spectra exhibit power law behavior for scales smaller than
  about 1.2\,Mm. 
}\label{fig:magkin}
\end{figure}

For the comparison between the magnetic and kinetic energy spectra in
Fig.~\ref{fig:magkin} we have chosen $\rho=10^{-5}$\,kg m$^{-3}$
(or $10^{-8}$\,g cm$^{-3}$, the units normally used in model atmosphere
tabulations), which is representative for the height of line
formation \citep[cf.][]{stenflo-shchukina01}. We find that for scales
smaller than about 2\,Mm the kinetic 
and magnetic energy spectra nearly coincide. This implies that there
is equipartition between the magnetic and kinetic energies at scales in
the range 0.3-2.0\,Mm.  Both spectra exhibit a steep
slope that can be described by a power law with exponent $-3.23$ for
scales smaller than about 1.2\,Mm. 

The equipartition between the magnetic and kinetic energies is of
course independent of the type of 1-D or 2-D representation that we
use for the spectral energy density, since the mathematical
expressions that relate the different representations with each other
are identical for the magnetic and velocity fields. 

Note however that the equipartition that we see in
Fig.~\ref{fig:magkin} refers to the conditions in the upper photosphere, at a height of
400-450\,km above the layer where the green continuum is formed. It
does not at all imply that there is equipartition at other heights. On
the contrary, since the density $\rho$ decreases almost exponentially
with height, while the magnetic and velocity fields vary much less, we
expect the kinetic energy density to dominate at the bottom of the
photosphere (and below) for the range of resolved scales that we are
considering. At the level of continuum formation, for instance, the
density is about 30 times the density in our line forming layer
\citep[cf.][]{stenflo-vernazza81}.

\section{Contribution of the flux tubes to the energy spectrum}\label{sec:fluxtubes}
In \citet{stenflo-s10aa,stenflo-s11aa} it was shown how the line-ratio
technique, when applied to the Hinode quiet-sun data set that we are
using here, reveals two distinct magnetic flux populations,
representing collapsed (kG) and uncollapsed fields. The collapsed flux
tube population was found to be the dominating contributor to the
measured circular polarization signals that exceed 0.5\,\%.  

The line-ratio information also allowed us to find statistical
relations between the vertical flux densities ${\bar B}_z$ and the
intrinsic field strengths of the flux tubes, and via the determined
flux tube filling factors to obtain statistical estimates of the flux
tube diameters $d_f$. The derived histograms for the flux tube sizes
showed that most of the flux tubes exist in the 10-70\,km size
range, well beyond the Hinode resolution limit, but within reach of
future telescope systems. It was also shown why this general size range can be
expected from the theory of flux tube collapse. 

The existence of a collapsed flux population implies that there is a concentration
of magnetic energy in the 
wave number range represented by the flux tubes. We therefore expect
the spectral energy density to have a significant bump at these
wave numbers. We will now try to model this spectral bump, based on the
observationally determined histogram of flux tube sizes from
\citet{stenflo-s11aa}. 

\begin{figure}
\resizebox{\hsize}{!}{\includegraphics{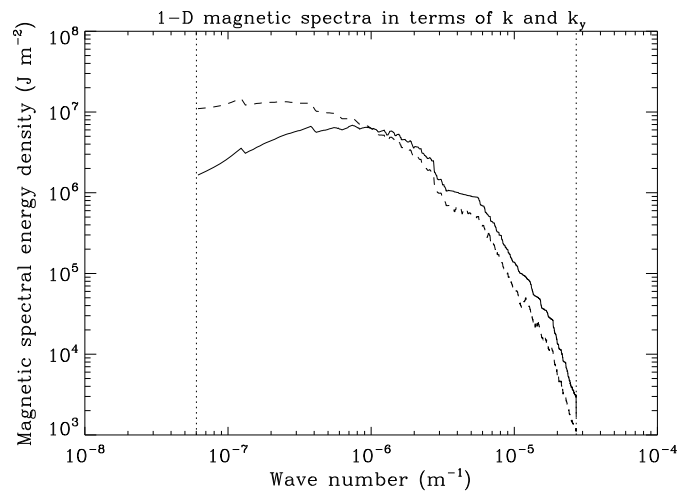}}
\caption{Comparison between the 1-D magnetic energy spectra ${\bar E}_{1D}(k)$
  (solid) and ${\bar E}_y(k_y)$ (dashed). They have the same slopes in the
  range with approximate power law behavior at 
  large wave numbers.  
}\label{fig:kspectrum}
\end{figure}

The theoretical modeling of the flux tube spectral contributions is
done in the $k$ wave number domain, not in the $k_y$ domain that we
used so far for the observationally determined spectra. As we will
later combine the observationally and theoretically determined
contributions, we need to convert the observational spectrum from
$k_y$ to $k$ space. We have done this with the help of Eqs.~(\ref{eq:12rel}) and
(\ref{eq:abel}). The application of Eq.~(\ref{eq:abel}) is numerically
tricky since it contains the derivative of the $k_y$ spectrum, which
needs to be smoothed to avoid large noise fluctuations. The result of
the conversion is shown as the solid line in Fig.~\ref{fig:kspectrum},
together with the original $k_y$ spectrum (dashed). 

To model the flux tube contribution it is sufficient for our purposes
to represent the magnetic field 
strength across a flux tube in the image plane with the following step
function: 
\begin{equation}
B(r)=B_f \,\,\Pi\left({r\over 2r_0}\right)\,,\label{eq:fthole}
\end{equation}
where $B_f$ is the intrinsic field strength of the flux tube, and the
function $\Pi(r/(2r_0))$ is unity for $r\le r_0$ and zero
otherwise. $2r_0$ is thus the diameter $d_f$ of the flux tube. 

The power spectrum of $B(r)$ is 
\begin{equation}
\vert{\tilde B}(k,d_f)\vert^2 \,\sim\, [B_f(d_f)]^2\,[J_1(k\,d_f/2)]^2\,/\,k\,,\label{eq:bpower}
\end{equation}
where the constant of proportionality, which is the same for the whole
flux tube population, will be determined later by normalization. $J_1$
is the Bessel function of order 1. The integral of this expression
over all $k$ is proportional to $d_f^2 B_f^2$, as it must be to
represent the energy contribution of a flux tube, since $d_f^2$ is
proportional to the filling factor, $B_f^2$ to the magnetic energy density of
a flux tube. Note
that the 2-D power spectrum ($E_{2D}(k)$) of the function $\Pi$ would
be the Airy function with $k^2$ in the denominator, but as we are
dealing with the 1-D spectrum ($E_{1D}(k)$), the denominator contains
the unsquared $k$. 

The power spectrum $\vert{\tilde B}(k,d_f)\vert^2$ is thus proportional to the
magnetic energy per unit wave number bin due to one flux tube of size
$d_f$. If we have many flux tubes, we simply add their power spectra
to obtain the combined contribution, since the spatial locations in
the image plane of the various flux tubes can be assumed to be
uncorrelated, without any systematic phase relations in the Fourier
domain. Therefore the superposition in $k$ space of the various flux
tubes has no coherency but can be done incoherently. 

In \citet{stenflo-s11aa} we determined the histogram $h_f(d_f)$ of the flux tube
sizes $d_f$ from the observational data. Since the determination
depended significantly on the relative contribution of the collapsed
flux population in the noise-dominated region with Stokes $V$
polarization amplitudes less than 0.5\,\%, two versions of the
histogram were derived, based on two different assumptions for the
contributions from the very small flux densities. 

To obtain the combined contributions to the energy spectrum from all
the flux tubes we need to multiply the power spectrum for a single
flux tube with the distribution function $h_f$ and integrate over
$d_f$. This integral can be written as the sum 
\begin{equation}
F(k)\,=\,F_0\,\sum_i\,\vert{\tilde B}(k,d_{f,i})\vert^2 \,h_f(d_{f,i})\,\Delta_i\,,\label{eq:fksum}
\end{equation}
where $F_0$ is a normalization factor, and we sum over all the $d_f$
bins, each having width $\Delta_i$ and representing flux
tubes of size $d_{f,i}$. The normalization factor is determined by the
requirement that 
\begin{equation}
\int F(k)\,{\rm d}k\,=\,{\cal E}_f\,,\label{eq:fnorm}
\end{equation}
where the average magnetic energy density contributed by the flux
tubes, ${\cal E}_f$, has been extracted from the analysis of the data
set in \citet{stenflo-s11aa} and therefore represents an empirical
constraint on the model. We use ${\cal E}_f=30.4$ in SI units, which
corresponds to an RMS field strength of 87\,G. Note that while the
flux tube field strengths $B_f$ are of order kG, their filling factors
are small, which together leads to an intermediate RMS value. 

\begin{figure}
\resizebox{\hsize}{!}{\includegraphics{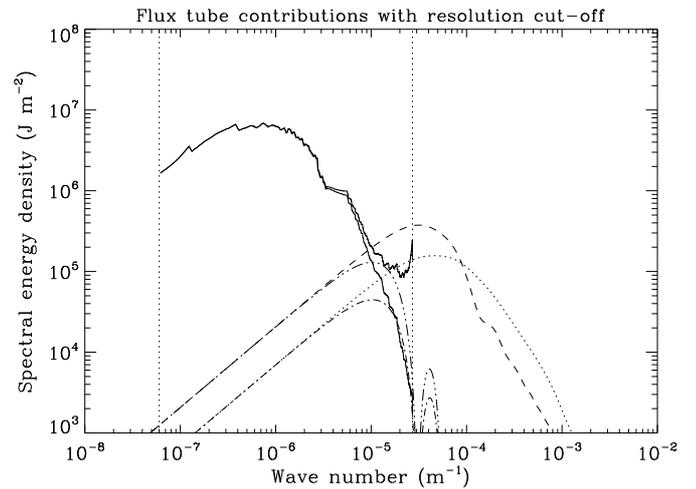}}
\caption{Modeled flux tube contributions to $E_{1D} (k)$ (dashed and dotted curves)
  based on two somewhat different versions of the flux tube
  distributions derived in \citet{stenflo-s10aa}. When the MTF of the
  Hinode telescope, $T(k)$, is applied, the two curves get truncated
  near the Hinode resolution limit and become the dash-triple-dot and
  dash-dotted curves. The lower of the two solid curves is identical
  to the solid curve in Fig.~\ref{fig:kspectrum}. When this curve is
  divided by $T(k)$ to compensate for the MTF quenching, the upper solid
  curve is obtained. 
}\label{fig:ftcont}
\end{figure}

The resulting, normalized $F(k)$ spectra for the two different
versions of the size histogram $h_f$ from
\citet{stenflo-s11aa} are shown as the dashed and dotted curves in
Fig.~\ref{fig:ftcont}, together with the observationally determined
${\bar E}_{1D}(k)$ spectrum from Fig.~\ref{fig:kspectrum} as the solid
line. 

Due to the finite resolution of the observations, the  $F(k)$ spectra
get truncated by the telescope function $T(k)$, the Hinode MTF, as in
Eq.~(\ref{eq:etk}). We let the MTF here be approximated by the Airy
function 
\begin{equation}
T(k)\,=4\,[\,J_1(0.5d_r\,k)\,/\,(0.5d_r\,k)\,]^2\,,\label{eq:tmtf}
\end{equation}
where the factor 4 makes $T(k)$ amplitude normalized to unity for
small wave numbers. In this representation, the spatial resolution is
given by an averaging circular area of diameter $d_r$. Since the
lower limit for the resolution element must be two pixels, which with the
present pixel size of 0.16\,arcsec is 232\,km (with 725\,km per
arcsec), we use the rounded value of $d_r=250$\,km for the effective
resolution when calculating $T(k)$. 

After application of the MTF, the $F(k)$ spectra get truncated into
the dash-triple-dot and dash-dot curves in
Fig.~\ref{fig:ftcont}. These truncated spectra are in their descending
branches quite similar to the observed spectrum ${\bar E}_{1D}(k)$,
indicating that the spectral shape in this range is largely
determined by the MTF and not by the shape of the intrinsic
spectrum. 

In principle one may retrieve the intrinsic spectrum $E_{1D}(k)$
through division of ${\bar E}_{1D}(k)$ by $T(k)$, although in practice
this produces errors that escalate towards infinity where $T(k)$ goes
to zero near the cut-off. Still such a division may give an indication of
the true shape of the spectrum. The result of this division is shown
by the upper solid curve in Fig.~\ref{fig:ftcont}. It demonstrates
that the previously determined steep power law exponent of $-3.23$ in
Fig.~\ref{fig:magkin} is not representative of the intrinsic spectrum,
which is much less steep. 

\begin{figure}
\resizebox{\hsize}{!}{\includegraphics{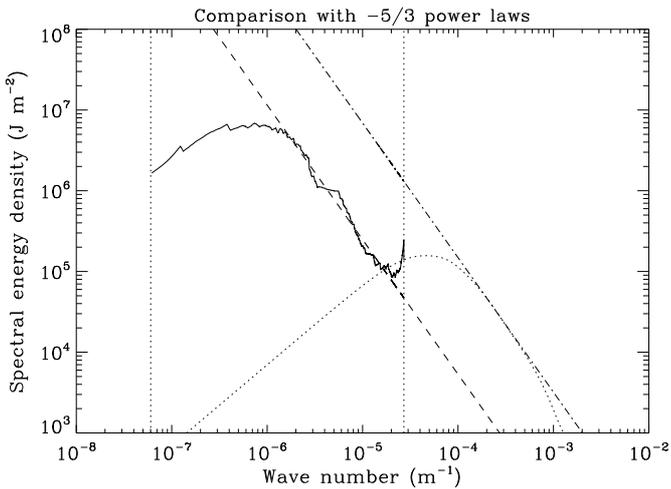}}
\caption{Comparison between the spectral shapes and power laws with exponent 
  $-5/3$. The solid and dotted curves are identical to the 
  corresponding curves in Fig.~\ref{fig:ftcont}. 
}\label{fig:53laws}
\end{figure}

When we instead compare the slopes of the corrected
observational spectrum and of the flux tube $F(k)$ spectra
with power laws, we find that power laws with exponent $-5/3$
provide an optimum fit, as demonstrated in Fig.~\ref{fig:53laws}
(where we have simplified the plot by showing only one of the two
versions of the $F(k)$ spectrum). \citet{stenflo-kolmogorov41} theory
explains in terms of  purely 
dimensional arguments how a $-5/3$ power law is produced when the
energy is injected at large scales and then cascades down 
the scale spectrum with a constant energy transfer rate 
per unit mass from smaller to larger wave numbers. The range of wave
numbers between the scale at which the energy is 
fed in and the cut-off scale at the diffusion limit is called the
inertial range, over which the $-5/3$ power law applies.

\section{Intrinsic magnetic energy spectrum}\label{sec:intrspec}
In the solar case the inertial range that we see at resolved scales
gets interrupted by the convective instability that causes the spontaneous
collapse of magnetic flux into flux tubes. The collapse process is
driven by the external gas pressure, which very effectively
injects energy into the magnetic field at the scales
that are populated by the flux tubes. This produces a significant
bump or enhancement of the magnetic energy spectrum at these scales. 

While the $-5/3$ power law seems to describe the descending branch of
the flux tube spectrum $F(k)$ over a substantial range, the dotted
curve in Fig.~\ref{fig:53laws} starts to drop off
much faster near $k=0.001$\,m$^{-1}$ (corresponding to a scale of
6\,km). This cut-off is however an artefact of the circumstance that
$F(k)$ exclusively represents the flux tubes and ignores the smaller-scale
fields beyond this range. The most likely scenario is that the
magnetic energy that has been fed into the flux tubes by the collapse
process later (as part of the flux tube decay process) cascades down
the scale spectrum until the magnetic diffusion 
limit is reached. In the absence of other scale-dependent mechanisms
between the flux tube and the diffusion scales, the most natural
assumption is that, as in Kolmogorov theory, the rate of energy
transfer from smaller to larger wave numbers stays constant. The
resulting energy spectrum will then be a $-5/3$ power law also over this intertial range. 

These considerations can now be used to construct an intrinsic magnetic
energy spectrum that is independent of any telescope resolution and
which covers all wave numbers, from the largest scales all the way down to
the magnetic diffusion cut-off. For the small wave numbers the
observed but MTF-corrected $E_{1D}(k)$ spectrum applies. When this spectrum
encounters the flux tube spectrum near the Hinode resolution limit,
the flux tube spectral contribution takes over. We then let the
descending branch beyond the flux tube maximum 
be continued as a $-5/3$ power law all the way down to the
diffusion cut-off. 

\begin{figure}
\resizebox{\hsize}{!}{\includegraphics{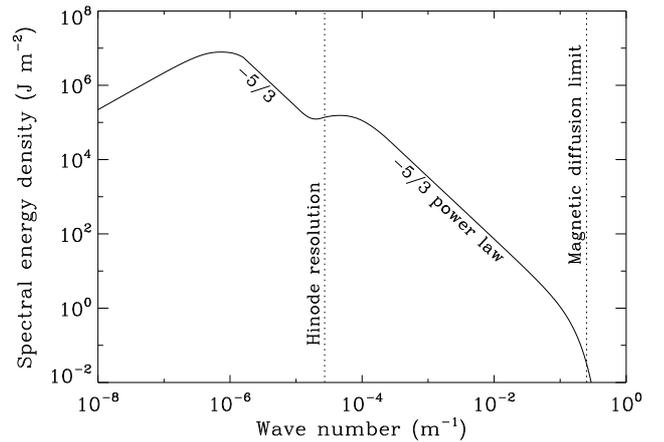}}
\caption{Overview of our best estimate for the quiet-sun magnetic energy spectrum
  $E_{1D} (k)$, down to the fundamental cut-off at the magnetic diffusion
  limit (located at a scale of 25\,m), where the magnetic field ceases
  to be frozen-in and decouples from the plasma. The spectral bump
  due to the collapsed flux tubes just beyond the Hinode resolution
  limit is surrounded by inertial ranges with $-5/3$ power law
  behavior. While the flux tubes and the flux at larger scales are
  oriented preferentially in the vertical direction, the field below
  the flux tube scale range will have a much wider and randomized angular
  distribution. These small scales contain the ``hidden'' flux that
  was revealed by the Hanle effect three decades ago.  
}\label{fig:overv}
\end{figure}

The resulting magnetic spectrum is shown in Fig.~\ref{fig:overv}. A
smooth representation of the resolved part has been achieved by
using an analytical $-5/3$ power law for the first inertial range, and
letting the bump around wave numbers $10^{-7}$ - $10^{-8}$ be given
the same shape as the bump described by the dotted curve in
Fig.~\ref{fig:53laws}. For convenience, we let the quenching of
the spectrum at the magnetic diffusion limit be described by a
function of the same analytical form as the function $T(k)$ of
Eq.~(\ref{eq:tmtf}), but inserting for  the ``resolution'' scale
$d_r$ the value 25\,m (rather than the 4 orders of magnitude larger
value of 250\,km for the Hinode resolution). 

The magnetic diffusion limit is the spatial scale where the field line 
diffusion time scale becomes equal to the convective time scale, in
other words, where the magnetic Reynolds number becomes unity. Below
the diffusion limit the field lines cease to be frozen-in and decouple
from the plasma. The scale where this happens was estimated in
\citet{stenflo-dewijn09} to be 15\,m, based on 
Kolmogorov theory, the Spitzer conductivity for a temperature of
10,000\,K, and observational constraints for the velocities. The
$10^4$\,K value was used, because much of the magnetic structuring
that we see throughout the photosphere is
generated by the turbulence just below the photosphere. Even if the
magnetic field decouples from the plasma earlier in the higher layers
of the photosphere (like in the line-forming layers), structures 
generated lower down where the frozen-in condition still applies will
be mapped by the field-line connectivity throughout the photosphere. The Spitzer
conductivity may overestimate the actual conductivity, in which case
the scale for the diffusion limit becomes larger. The estimate of
25\,m that we here adopt for the diffusion limit should be subject to
improvement in future work.  

\begin{figure}
\resizebox{\hsize}{!}{\includegraphics{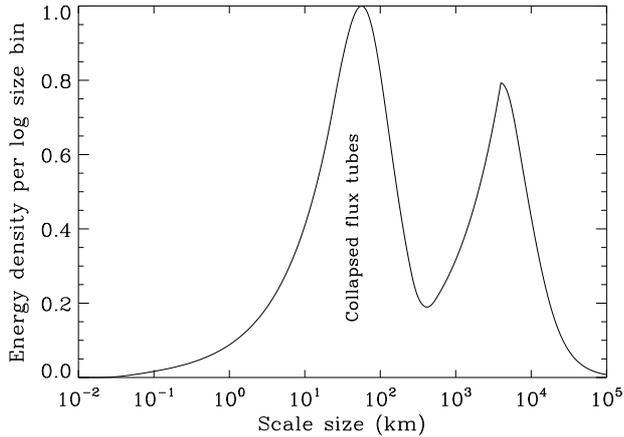}}
\caption{Conversion of the $E_{1D} (k)$ spectrum in Fig.~\ref{fig:overv}
  to a histogram that represents the relative amount of magnetic
  energy in bins of $\log d$, where the scale size $d$ is defined by
  $d=2\pi/k$. The two bumps in the $E_{1D} (k)$ spectrum now become more
  pronounced, in particular since we use a linear vertical scale, and
  the $-5/3$ power laws get converted into $d^{2/3}$ 
  functions that govern the shapes of the left sides of the two
  peaks. 
}\label{fig:ensize}
\end{figure}

The spectral energy density $E_{1D} (k)$, which represents the
magnetic energy per unit wave number, can easily be converted into a
spectral energy density $E_{\rm size}^\prime(d)$ representing the energy per unit
bin in $\log d$, where $d$ is the scale size, related to the wave
number through $d=2\pi/k$. As 
\begin{equation}
E_{\rm size}^\prime(d)\,{\rm d}\log d\,=\,E_{1D} (k)\,{\rm d}k\,,\label{eq:esize}
\end{equation}
we obtain 
\begin{equation}
E_{\rm size}^\prime(d)\,=\,(\ln 10)\,k\,E_{1D} (k)\,.\label{eq:edk}
\end{equation}
This relation is used to convert the spectrum in Fig.~\ref{fig:overv}
to the energy size distribution (with amplitude normalization) in
Fig.~\ref{fig:ensize}. 

The two bumps in the $E_{1D} (k)$ spectrum appear much more pronouced
in Fig.~\ref{fig:ensize} since we now use a linear instead of logarithmic
vertical scale. The $k^{-5/3}$ power laws get converted into $d^{2/3}$
power laws that determine the shapes of the left wings of the two
peaks in Fig.~\ref{fig:ensize}. The collapsed flux tubes dominate the
magnetic energy contribution over the approximate range
10-150\,km. Note that the peak representing the flux tubes is shifted 
towards somewhat larger sizes than the 
histogram of flux tube sizes in \citet{stenflo-s11aa}, which was most 
prominent in the range 10-70\,km. The reason for this is that
the flux tube field strength $B_f$  increases with size, and for the
energy contributions in Fig.~\ref{fig:ensize} the size distribution gets
weighted with $B_f^2$, which significantly enhances the contributions
from the larger scales. 

Since the power law behavior for the small scales (below about 10\,km)
implies that the field strength scales as $B\sim d^{1/3}$ (as the
magnetic energy scales with $ d^{2/3}$), the fields at a
scale of for instance 80\,m are about 5 times weaker than at the 10\,km
scale. Although this is a rather slow decline of field strength with
decreasing scale size, it is significant and indicates that the main
contribution to the observed Hanle depolarization  caused by the
``hidden'' flux of tangled fields probably comes from fields with
sizes of order 1\,km. Much smaller sizes are less effective in
producing the Hanle signature due to their weaker fields, while much larger
sizes also contribute less, because they are dominated by the flux
tubes, which are preferentially vertically oriented and for this reason do
not contribute to the Hanle effect. We need to get down to scales
where the angular distribution of the field vectors has become
sufficiently wide, while the field has still retained enough strength. The Hanle
effect and the tangled field will be explored more in the next section.

\section{Probability density function for the vector field}\label{sec:pdf}
While the energy spectra provide information on the sizes of the flux
elements, weighted by the square of the field strength, there is no
direct information on the distribution of field strengths or flux
densities, for which we instead need the probability density
function (PDF). Although the PDF contains no direct information on the
size distribution, it is strongly influenced by it and is in
particular a very sensitive function of the telescope resolution
cut-off. 

There are two main effects of the finite resolution: (1) The
contributions from subresolution elements, like flux tubes, get
reduced because their filling factors are smaller than unity. (2) The
topological mixing of opposite magnetic polarities leads to
cancellation of the contributions to the circular polarization from the
Zeeman effect, making such mixed-polarity flux invisible in
magnetograms. In contrast to the Zeeman effect, such cancellation does
not apply to the Hanle effect, since the 
depolarization effect does not change sign when the
orientation of the field is reversed. 

Due to the small-scale structuring, the quiet-sun field that has been
smoothed by the resolution element is very different from the
unsmoothed field. To avoid confusion we therefore reserve the term
``field strength'' for the field as seen with infinite resolution (the
intrinsic, resolution-independent field), while the corresponding term
for the smoothed field is ``flux density''. While the flux densities
represent the observed quantities, the field strengths can only be inferred
indirectly with the help of idealized models. 

\begin{figure}
\resizebox{\hsize}{!}{\includegraphics{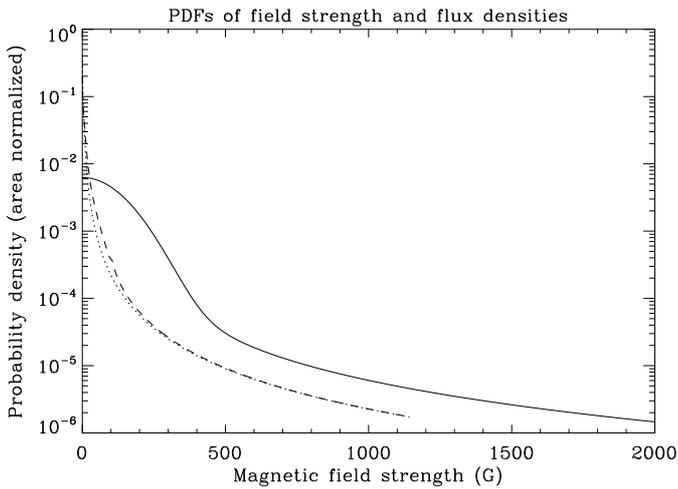}}
\caption{Comparison between the area-normalized probability distribution functions
  (PDF) for the field strengths (solid line) and the flux densities as
  observed with the Hinode resolution (dotted line for the vertical
  flux densities, dashed line for the total flux densities). The
  extended wings that decline quadratically are mainly due to the
  collapsed flux tubes. The large width of the Gaussian
  core region for the field strength PDF is required to
  satisfy the observational constraints from the Hanle effect. Due to
  cancellation of mixed polarities this Gaussian core contracts to
  become the narrow stretched exponential core for the flux density
  PDFs. 
}\label{fig:pintr}
\end{figure}

The distinction between these two concepts is illustrated in
Fig.~\ref{fig:pintr}, where we compare the PDFs for the observed flux densities
of our Hinode quiet-sun data set (dotted curve for the
vertical flux densities, dashed curve for the total flux densities)
with the inferred PDF for the intrinsic field strengths (solid
curve). The procedure by which the properties of this field-strength PDF
have been inferred and constrained by observations will be described next. 

Let us however start by first considering the known properties of the PDF
for the flux densities. As shown in \citet{stenflo-s10aa}, this PDF can be
closely fitted by an analytical function that is represented by a
stretched exponential in the extremely peaked core region, with the
peak centered at
zero flux density. The PDF has extended wings that decline quadratically with
flux density. While the extended wings get their contributions from
the largest (and strongest) flux tubes, the core region contains a
mixture of contributions from the smallest flux tubes and from
uncollapsed fields. ``Hidden'' flux for which the polarity
contributions cancel within the resolution elements do not contribute
anywhere at all to the PDF for the flux densities, although their
existence is revealed by the Hanle effect \citep{stenflo-s82}. 

The cancellations do not much affect the outer wing regions of
the PDF, since the low probability density there implies that the
contributing larger flux tubes are relatively rare and therefore
sparsely spaced with respect to each other. Instead, the resolution
effect is mainly a filling-factor effect: the flux densities are smaller
than the intrinsic field strengths in proportion to the filling
factor. This leads to a contraction of the
field-strength scale to generate a flux density scale, which
approximately preserves the general shape of the PDF wings but
compresses them to smaller field values. To retrieve the wing PDF for the
intrinsic field strengths we need to divide the flux density values by the filling factor. 

The determination of the wing region of the solid curve in
Fig.~\ref{fig:pintr} has been based on these considerations. The
maximum flux density measured in our data set is 1145\,G, while the
line-ratio information implies that this is due to a flux tube with
intrinsic field strength 1840\,G and filling factor 
62\,\%. It is the largest filling factor found in this particular data set,
but it is still relatively far from 100\,\%.  The value 1145\,G
represents the end point of the flux density PDFs in
Fig.~\ref{fig:pintr}, and stretched in the horizontal
direction by the factor 1/0.62 brings it to the solid curve and the 
value 1840\,G. Other portions of the wing curve can be regarded as
representing similar scalings.   

The properties of the core region of the intrinsic PDF is governed by
entirely different considerations, namely by the observational
constraints from the Hanle effect. It was realized long ago
\citep{stenflo-s82,stenflo-s87} that there could only be one consistent
explanation for the observed deficit in the scattering polarization that
is theoretically expected, namely depolarization by magnetic fields
(Hanle effect) that are ``hidden'' in the sense that they do not show
up in solar magnetograms due to spatial mixing of the magnetic
polarities on scales much smaller than the telescope resolution. It
was concluded 
that this hidden field needs to have a strength in the range
10-100\,G to be consistent with the observations. The many papers
since then on the theoretical development of the theory of polarized line
formation with partial frequency redistribution in magnetized media
together with breakthroughs in the observational techniques to record
the Second Solar Spectrum (the linearly polarized spectrum that is
formed by coherent scattering processes, and which is the playground
for the Hanle effect) have confirmed the validity of these early
insights and led to refined constraints on the hidden field. 

The most detailed modeling so far of the Hanle effect produced by the
hidden field \citep{stenflo-trujetal04} used 3-D atmospheres generated
by numerical simulations of magnetoconvection to model the scattering
polarization that has been observed in the Sr\,{\sc i} 4607\,\AA\ line
and in molecular C$_2$ lines. While the Sr line requires a 60\,G field
to fit the observations, a much weaker field of about 10\,G is needed
for the C$_2$ lines. \citet{stenflo-trujetal04} suggested that
  the explanation for this discrepancy by a factor of 6 lies in a
  correlation between the hidden magnetic flux and the solar
  granulation. The C$_2$ lines are formed almost entirely in the
  interior of the granulation cells, where the hidden field is weaker,
  while the formation of the Sr line gets significant contributions
  from the intergranular lanes, where the hidden field is much stronger.  

This explanation is supported by the finding in \citet{stenflo-s11aa}
that the collapsed flux population is preferentially located in the
intergranular lanes, because the
magnetic energy that has been injected into the magnetoturbulent 
spectrum by the flux collapse
process cascades down the scale spectrum when the flux tubes dissolve
(fraying by the interchange instability) and is thus directly fed into
the inertial branch where the hidden flux resides. This physical connection
between flux tubes and hidden flux implies that they should 
follow the same spatial distribution. 

While the hidden flux is thus expected to be spatially structured, we
will in the following restrict the discussion to the spatially averaged constraint
represented by the 60\,G value for the Sr line. Note however
that the Hanle field strengths that we have quoted 
so far have all been based on an interpretational model, according to
which the hidden flux consists of optically thin elements that have an isotropic angular
distribution and a single field strength, i.e., a 
PDF in the form of a $\delta$ function. In contrast, the real Sun should be
characterized by continuous PDFs, which get constrained by the
observed net depolarization effect. The quantitative details will
depend on the shape of the PDF. 

As will be shown below, the extended PDF wings do not play any significant role for the
observed Hanle depolarization effect. Instead nearly the entire effect must 
have its source in the PDF core region. The field-strength width of this
core region must be about two orders of magnitude larger than the
observed PDF core for the flux densities, which can be fitted with a
stretched exponential that has a half width of only 1.2\,G. No
stretched exponential function can be used for the field-strength PDF if it is to
satisfy the Hanle contraints. 

It is natural to expect that the ``hidden'' fields in the inertial range below
the flux tube scales would get randomized by the
small-scale turbulence, as the scales are much smaller than the pressure scale
height, which implies that the turbulence becomes nearly isotropic. This favors the
development of a Gaussian field distribution. These considerations lead us
to the choice of a Gaussian shape for the very wide PDF core region
that represents the intrinsic field strengths. 

The combination of a Gaussian core and quadratically declining
extended wings is satisfied by a Voigt function $H(a,b)$, where $a$ is
the damping parameter, and $b=B/B_D$ is the dimensionless field strength,
normalized by the analog of a ``Doppler'' width $B_D$. The function
given as the solid line in 
Fig.~\ref{fig:pintr} has $a=0.053$ and $B_D=170$\,G. We further assume 
that the PDF ends abruptly at 
2000\,G, since stronger fields can hardly be contained by the photospheric
gas pressure, and larger structures like sunspots that have stronger
fields are not representative of quiet solar regions. The one-sided
truncated Voigt function in Fig.~\ref{fig:pintr} has been normalized to unit
area. It has a half width of 146\,G. While the true solar PDF may have
a shape that differs from a Voigt function, it must have a core with a
similar large half width to satisfy the Sr\,{\sc i} Hanle constraint, 
and have extended wings to satisfy the line-ratio constraints. Therefore
the true general appearance of the PDF cannot differ that much from our chosen
Voigt function. 

\begin{figure}
\resizebox{\hsize}{!}{\includegraphics{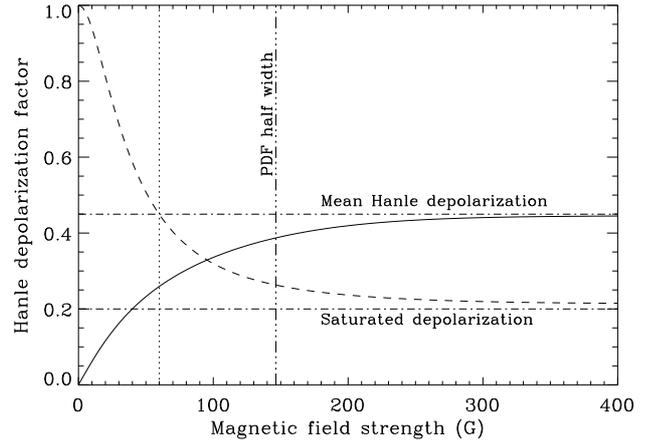}}
\caption{PDF contributions to the mean Hanle depolarization of 0.45
  observed in the Sr\,{\sc i} 4607\,\AA\ line, assuming an isotropic
  angular distribution. The dashed curve represents the Hanle
  depolarization factor as a function of field strength. The value
  0.45 is reached for 60\,G, which would be the field strength of a
  single-valued PDF. The solid curve is the cumulative contribution to
  the Hanle depolarization from the PDF for all fields weaker than the
  given field strength $B$. The bulk of the contribution comes from
  the fields weaker than the PDF half width (146\,G). 
}\label{fig:hanle}
\end{figure}

The way in which the different parts of the PDF contribute to the
observed Hanle depolarization is illustrated in
Fig.~\ref{fig:hanle}, where the Hanle depolarization factor $k_H$ is
plotted as a function of field strength $B$ as the dashed curve. It starts
from unity (implying no depolarization) for zero field, and decreases
to asymptotically approach the saturation limit of 0.2 (implying
80\,\%\ depolarization) as the field strength goes to infinity. The
value of the saturation limit depends on the assumed angular
distribution of the field vectors and is 0.2 for the isotropic case
\citep{stenflo-s82,stenflo-s87}. The observational constraint is a depolarization
factor of 0.45. This value is reached by the dashed line for 60\,G, the field
strength of  a $\delta$ function PDF that would satisfy the Hanle
constraint. 

Let us use $P(B)$ to denote the
area-normalized version of our Voigt function. The net Hanle depolarization factor 
is then obtained from integration over the product of $P$ and $k_H$. To
illustrate how this integral builds up as we go from the inner to the
outer parts of the PDF we have in Fig.~\ref{fig:hanle} plotted as the
solid curve the cumulative depolarization factor $C_H(B)$, defined as 
\begin{equation}
C_H(B)=\int_0^B P(B)\,k_H(B)\,{\rm d}B\,.\label{eq:ch}
\end{equation}
In the limit of large $B$ the solid line reaches the mean Hanle
depolarization factor of 0.45, as required. We see that most of the
contribution comes from the core region inside the core half
width (146\,G), while there is no significant contribution from the
wings beyond about 250\,G. This is another way of saying that the
collapsed flux tubes are not relevant for the observed Hanle effect.

\section{Energy and sizes of the hidden, tangled field}\label{sec:hidden}
A long-standing question has been whether the stored magnetic energy
of the tangled, hidden flux is significant for the energy balance of
the solar atmosphere \citep[cf.][]{stenflo-trujetal04}. We know that
the magnetic energy stored in kG-type flux tubes is comparable to the
ambient kinetic energy, since it follows from the lateral pressure
balance that the magnetic pressure is comparable to the ambient gas
pressure (otherwise kG fields of narrow flux tubes could not be contained in the
photosphere). In the following we will show that the energy in the
hidden magnetic flux is comparable to the average energy that is due to the
flux tubes. This implies that the hidden flux is as 
relevant to the energy balance of the solar atmosphere as the flux
tubes are. 

\begin{figure}
\resizebox{\hsize}{!}{\includegraphics{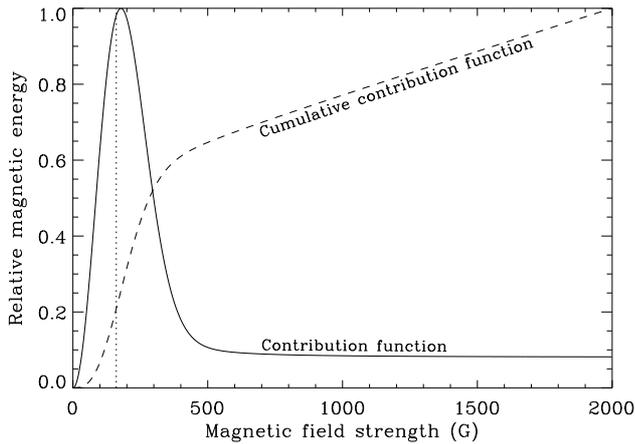}}
\caption{Relative contributions to the total magnetic energy from the
  different parts of the field strength PDF. While the solid line is
  the amplitude-normalized contribution function, the dashed line
  represents its integral from zero up to the given field strength
  (the cumulative contribution function), normalized to unity
  amplitude. We notice that the Gaussian PDF core is responsible for
  approximately 60\,\%\ of the total energy, while the extended damping wings
  (due to the flux tubes) contribute the rest. The vertical
  dotted line marks the RMS ($\sqrt{\langle B^2\rangle}$) of the field,
  161\,G. 
}\label{fig:endist}
\end{figure}

We address this question here by illustrating in
Fig.~\ref{fig:endist} the relative contributions to the total
magnetic energy that come from the different parts of the PDF $P(B)$ for the
intrinsic field strengths $B$. For reference we mark the RMS value of
the field ($\sqrt{\langle B^2\rangle}$) by the vertical dotted
line. The solid curve that is labeled 
``contribution function'' is $P(B)\,B^2$, normalized to its maximum
value. The peak around 200\,G is produced by the Gaussian core of the PDF,
while the PDF wing is responsible for the elevated ledge that stays
constant until the PDF ends, since the multiplication of the
quadratically declining $P(B)$ with the quadratic function $B^2$ makes
a constant product. 

To compare the relative contributions of the Gaussian core, which is
the domain of the hidden flux, and the PDF wings, which is the domain
of the collapsed flux tubes, we compute the cumulative contribution
function $C_E(B)$, defined as 
\begin{equation}
C_E(B)=\int_0^BP(B)\,B^2\,{\rm d}B\,.\label{eq:ce}
\end{equation}
$C_E(B)$, after having been normalized to its maximum value, is
plotted as the dashed curve in Fig.~\ref{fig:endist}. It is
characterized by an initial steep rise, due to the Gaussian PDF core,
followed by a shallower straight line region due to the PDF wing. The
transition between these two regions occurs around a relative magnetic
energy of 0.6. This implies that about 60\,\%\ of the total magnetic
energy comes from the hidden, tangled field, while the remaining part
comes from the collapsed flux tubes. We can therefore conclude that
the hidden magnetic flux is as relevant to the overall dynamics and
energetics of the solar atmosphere as the flux tubes are.

\subsection{Role of the cancellation function}\label{sec:canc}
While the intrinsic PDF has an RMS field of 161\,G, the average field
(obtained from the integral of $P(B)\,B$) is 110\,G. In contrast to the
magnetic energy, the cumulative contribution function for
the flux has its source almost exclusively from the Gaussian PDF core, because
the contribution function for the flux does not have extended wings. 
Thus, while the magnetic energy is shared between the flux tubes and
the hidden flux in approximately equal proportions, the contribution
to the total flux is dominated by the hidden field. 

The average field strength of 110\,G is more than 5 times the average
flux density of 20\,G found for our Hinode data set
\citep{stenflo-s11aa}. This implies that about 80\,\%\ of the total
flux remains invisible at the Hinode resolution. The photosphere is a
seething ocean of tangled fields, and an instrument like Hinode is
able to see only the ``tips of the icebergs''. 

\citet{stenflo-pietarila09} used the cancellation function concept to
estimate how much intrinsic magnetic flux remains invisible to Hinode
due to cancellation of mixed polarities within the resolution element
and also arrived at the conclusion that at least 80\,\%\ of the total
flux is missed by Hinode. The cancellation function represents the
average unsigned flux density as a function of the spatial
resolution, i.e., the size $d$ of the smoothing window. Using the Hinode
data set one can derive this function by artificially smoothing the
data with running windows of various size. A power law $d^{-\chi}$ is
then found, characterized by the cancellation exponent $\chi$. 
If $\chi$ can be considered independent of $d$, this power law may be
extrapolated into the scale range below the resolution 
limit, to estimate what the average unsigned flux density would be if we
could resolve these small scales. 

Such extrapolation implicitly assumes that the field is scale
invariant over the extrapolation range. However, this assumption now
turns out not to be valid, since the instability process of flux  collapse breaks the
scale invariance and introduces, as we have seen, a preferred scale
around 10-100\,km, which is populated by the collapsed flux
tubes. 

The collapse process is the origin of the two observed, distinct populations: collapsed and
uncollapsed flux. The two populations can be expected to contribute to 
very different cancellation
functions. Since the flux tubes represent concentrated flux islands
that are generally well separated from each other (since they are
associated with small filling factors), there will be little mixing of
opposite polarities from adjacent flux tubes within the resolution
element. This implies a cancellation exponent $\chi$ that should be
small, near zero. In contrast, the uncollapsed flux is expected to be
rather randomized by the turbulent motions, leading to abundant mixing
and a large cancellation exponent. For a completely random field,
where the different spatial points are uncorrelated, $\chi$ would be
unity (since the relative fluctuations scale with $1/\sqrt{N}$, where $N$ is the
number of magnetic elements within a smoothing window, and $N$ is
proportional to window area $d^2$). 

While the half width of our PDF $P(B)$ for the intrinsic field
strengths is 146\,G, the half width of the PDF for the Hinode
absolute flux densities is only 1.2\,G. The PDF core region has
therefore become 
contracted by a factor of about 120 when going from the intrinsic
scales to the Hinode scale, due to flux cancellation within the
smoothing window. If we would apply the maximum possible cancellation
exponent of unity, treating the uncollapsed field as entirely random
without any spatial correlations, then we would get down to a scale size
for the ``intrinsic magnetic elements'' that is 120 times smaller than
the Hinode resolution scale, i.e., 2\,km. It may be interpreted as representing the largest
possible size of the magnetic elements that are responsible for the
broad Gaussian core of our PDF function $P(B)$. 2\,km is at the bottom of the
flux tube size range and near the top of the inertial range where the
hidden flux resides. Since the field strength $B$ is expected to
scale like $d^{1/3}$ over this inertial range, the strongest
fields will be located in the large-scale part of this range. It is the
strongest fields with a broad angular distribution that
contribute most to the observed Hanle depolarization signal. 

There is no direct connection between the cancellation function and
the scaling of the field strengths, since the cancellation function
depends on the topological mixing of the positive and negative
polarities. The cancellation function will not vary with scale size in monopolar regions,
while the field strength scale variations do not depend on the
polarity distributions. The field strength is expected to
decrease as we go down in scale in the inertial range, while the
cancellation function can only monotonically increase when the smoothing window
is decreased. 

The Hanle constraint depends on the intrinsic field strengths
and angular distributions, but not on the polarity mixing. Therefore
it is not the scaling of the cancellation function that is relevant
for the Hanle constraint (as incorrectly assumed in
\citet{stenflo-s11aa}), but the field strength scaling. While the 
angular distribution needs to be wide, since the Hanle effect is most
sensitive to horizontal fields but insensitive to vertical fields, the
polarities do not need to be mixed to generate Hanle 
depolarization. Wide angular distributions are possible also in
monopolar regions.

\section{Discussion}\label{sec:concl}
The magnetic energy spectrum extends over approximately 7 orders of
magnitude (from the magnetic diffusion limit of order 25 \,m,
to global scales of order 250\,Mm), while current telescopes can only
resolve 3 of these (for Hinode down to the resolution limit of about
250\,km). Besides the resolution cut-off the observationally
determined scale spectra in quiet solar regions are very sensitive to measurement noise,
which has a flattening effect. Therefore we have chosen to work with the
disk center quiet-sun data set that provides the optimum 
combination of low noise and high spatial resolution, obtained with
the Hinode SOT/SP instrument in deep integration mode. 

We find approximate equipartition between the observationally
determined kinetic and magnetic energy spectra over the
scale range 0.3-1.2\,Mm. This equipartition may be more accidental 
than it may first seem and be limited to the particular layer where
the Fe\,{\sc i} lines used for the Zeeman-effect analysis are formed,
about 400\,km above the continuum formation layer at the bottom of the
photosphere. Since the height variation of the mass density is much
steeper than the variation of the magnetic and velocity fields in the highly
stratified solar atmosphere, equipartition at the 400\,km level would
imply dominance of the kinetic energy at the bottom of the
photosphere, where magnetoconvection is more effective in producing
magnetic structuring. 

In idealized isotropic turbulence we may expect approximate scale
invariance, and as a consequence have a
cancellation function with a constant cancellation exponent that is associated
with a scale-invariant fractal dimension. The solar atmosphere is
however quite different with its pronounced stratification and
large superadiabaticity at the top of the convection zone, conditions which lead
to the instability of flux collapse. The circumstance that the collapse mechanism
requires the collapsing flux regions to be optically thick in the
horizontal direction (in the superadiabatic region, where the collapse
is driven) implies that there is a preferred scale size (the photon
mean free path in the superadiabatic layer), which breaks the scale
invariance.  The scale at which the invariance gets broken lies
just beyond the resolution limit of current telescopes but within
reach of the next generation of instruments. 

The collapse is driven by the external gas pressure, which converts
kinetic energy into magnetic energy in the form of the kG-type flux
tubes, creating a bump in the energy spectrum around the 10-100\,km
scales. The energy injected into this bump is expected to cascade down
the scale spectrum when the flux tubes decay via the interchange
instability. As observationally confirmed, the flux tubes are on
average nearly vertically oriented, 
both because they tend to be formed that way by adiabatic
downdrafts, and because the powerful buoyancy forces will try to make the anchored
field lines of the strong fields stand up. When the flux tubes dissolve into
weaker fields, their field lines will begin to be
tangled up by the turbulent motions. The resulting widening of the angular distribution of
the field vectors will make these fields effective in
depolarizing the scattering polarization and producing the
observed Hanle-effect signature that is the evidence for the real
existence of these ``hidden'' fields. 

In this scenario the magnetic energy spectrum of the hidden flux is
fed from the decay of the flux tubes, which implies that the
flux tubes and the tangled, hidden flux are physically related. They 
should therefore statistically be spatially distributed in similar
ways. It is known that the collapsed flux population is preferentially
located in the intergranular lanes, for instance from
  high-resolution imaging of G-band bright points
  \citep[e.g.][]{stenflo-berger_title96} as proxies for the
  concentrated magnetic flux, and recently from Stokes $V$ line-ratio
  observations with Hinode \citep{stenflo-s11aa}. It then follows that the hidden
field must also be significantly stronger in the intergranular lanes than
in the cell interiors. This provides an explanation for the finding
of \citet{stenflo-trujetal04} that the hidden field, when diagnosed
with the Hanle effect in the molecular C$_2$ lines, is about 6 times
weaker than the field strength found when using the Sr\,{\sc i}
4607\,\AA\ line. The C$_2$ lines are formed almost exclusively in the
cell interiors, where the hidden field is much weaker since it is fed
by the flux tube field, while
the Sr line can be assumed to represent the average atmosphere. As the Sr
line thus gets 
contributions from both the cell interiors and the intergranular
lanes, the hidden field would be found to be considerably 
stronger than suggested by our PDF for the intrinsic field strengths, 
if we could isolate the contributions from the intergranular
lanes. This conclusion is observationally supported by
\citet{stenflo-snik10}, who observed scattering polarization in the
molecular CN band with high spatial resolution that allowed the
intergranular lanes to be distinguished from the cell interiors. They
found the polarization to be significantly more suppressed in the
intergranular lanes, indicating that the hidden field is stronger
there. 

Our derived PDF for the intrinsic field strengths implies that there
is comparable amounts of magnetic energy in the flux tubes and in the
hidden field, which makes sense if the hidden field is fed from the
decaying flux tubes. The processes that drive the flux tube formation
and their dissolution take place much below the formation layer of the
Fe lines that we have used to diagnose the field. As the second inertial range at 
scales below 10\,km is much elevated because it has been fed by the
decaying flux tubes, there may be a dominance of the magnetic over
the kinetic energy densities in the line-forming layer 
at these scales. To be  physically significant the comparison
should however not be done for the line-forming layer, but rather for the lower layer
that plays the main role for the structuring that is diagnosed higher
up. The steep increase of the mass density
with depth elevates the kinetic energy spectrum to levels sufficient
to match the inferred small-scale magnetic energy spectrum.

\section{Concluding remarks}\label{sec:rem}
The overall magnetic energy spectrum that we have illustrated in
Fig.~\ref{fig:overv} is produced and sustained by convective
turbulence in a highly stratified medium. The term ``local dynamo'' is
often used to characterize the small-scale origin of the magnetic
structuring. However, as the scale spectrum continuously connects the
smallest with the largest scales over about 7 orders of magnitude,
there is no clear separation of scales as would be implied by the
terminology ``local'' vs. ``global''. The only preferred scale that
breaks the scale invariance seems to be the flux tube scale, which
separates the ``large'' scales (above 100\,km) from the ``small''
scales (below 10\,km). However, the term ``local dynamo'' is usually
not used to refer to this scale separation. Generally the term
``global dynamo'' is reserved for dynamo processes that have to do with the
hemispheric left-right symmetry breaking in cyclonic turbulence that
expresses itself in Joy's law for the tilt of bipolar magnetic regions
and in Hale's polarity law. However, as shown in \citet{stenflo-sk12},
Joy's law continues to be valid for small bipolar magnetic regions at
least down to scales that are covered by the upper portion of the
scale spectrum in our present analysis of Hinode
data. Therefore the magnetoturbulent spectrum that we have been
exploring here is relevant for the regeneration of the global solar 
magnetic field and for maintaining the 11-year cycle. Since it
does not make physical sense here to distinguish between a
``local'' and a ``global'' dynamo, we have avoided the use of such
dynamo terminology, and instead refer to the self-sustaining processes that maintain
the magnetoconvective spectrum in the stratified Sun. 

Most of the magnetic spectrum that we have derived resides at scales
that lie beyond the resolution of current telescopes. Therefore it has
been necessary to infer its shape from indirect observational
constraints, combined with general physical arguments. Since this
procedure is based on model assumptions
that may be questioned, our conclusions need to be validated by future
observational tests. While the predicted bump in the magnetic
energy spectrum lies at scales not yet covered by direct observations,
these scales are already covered by numerical simulations of
magnetoconvection. If the magnetic energy spectra produced by the
simulations do not show a bump or elevation of the spectrum
around the 10-100\,km scale, then either the simulations do not
include the needed physical ingredients and need to be overhauled, or
our present data analysis and modeling has been faulty.  Fortunately observational tests
that can settle this issue are almost around the
corner, since the critical scales lie just beyond the Hinode resolution limit, where we expect
the energy spectrum to get elevated due to the flux collapse process. It will
therefore not be long before this fundamental aspect of solar magnetism can
be either verified or refuted.

\begin{acknowledgements}
I want to acknowledge the fruitful, in-depth discussions about related issues that
took place at ISSI (International Space Science Institute) in Bern
during a meeting with an ISSI International Team, November 21-25,
2011. The analysis has been based on
observations done by the Hinode satellite. 
Hinode is a Japanese mission developed and launched by ISAS/JAXA, with NAOJ as domestic partner and NASA and STFC (UK) as international partners. It is operated by these agencies in co-operation with ESA and NSC (Norway).  
\end{acknowledgements}

\end{document}